# Tackling CS education in K-12: Implementing a Google CS4HS Grant Program in a Rural Underserved Area

Sherri Harms
Department of Computer Science and Information Technology
University of Nebraska at Kearney (UNK)
Kearney, NE 68849
harmssk@unk.edu

## Abstract

Providing computer science (CS) offerings in the K-12 education system is often limited by the lack of experienced teachers, especially in small or rural underserved school districts. By helping teachers in underserved areas develop CS curriculum and helping them become certified to teach CS courses, more young people in underserved areas are aware of IT-career opportunities, and prepared for CS education at the university level, which ultimately helps tackle the IT workforce deficit in the United States.

This paper discusses a successful implementation of a Google CS4HS grant to a rural underserved area, as well as lessons learned through the implementation of the program. Key elements in the implementation included a face-to-face hands-on workshop, followed by a seven week graduate-level online summer course for the teachers to learn and develop curriculum that covers the CS concepts they will be teaching. The teachers were supported with an online community of practice for the year as they implemented the curriculum.

## 1 Introduction

The Google CS4HS grant program [1] is dedicated to funding professional development opportunities for teachers. As stated on the Google CS4HS website, "Teachers are the foundation of excellence in computer science (CS). The world needs well prepared educators to facilitate CS not only because of the growing number of computing related jobs, but also because it develops critical thinking skills needed to solve complex problems, creativity that fosters new ideas, and skills to drive innovation in tech and other fields."

According to the CS4HS website, "Google launched CS4HS as an official program in 2009 to provide CS teachers globally with an opportunity to improve their technical and pedagogical skills." Today, they fund projects in over 20 countries. Figure 1 shows existing CS4HS programs in the United States. "CS4HS funding enables CS education experts to provide exemplary CS professional development for teachers. The funding focuses on three major growth areas for teacher PD in CS: 1. Facilitating the development and delivery of content that increases teachers' knowledge of computer science and computational thinking; 2. Allowing providers to customize learning content to meet local needs and the sharing of best practices for engaging all students;

and 3. Addressing the building of communities of practice that continue to support teacher learning throughout the school year." Currently, the CS4HS program funds are limited to $35,000 per year.

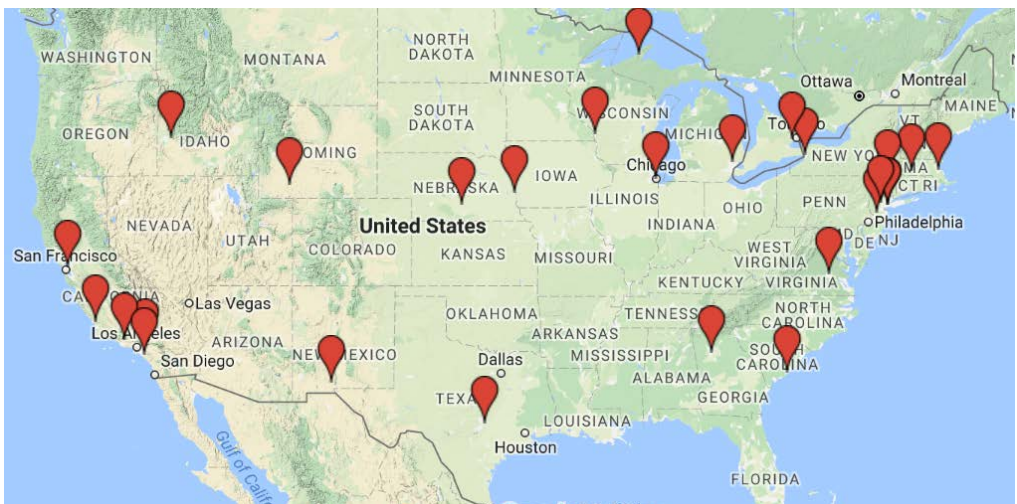

*Figure 1. Current GoogleCS4HS sites in the U.S.*

## 2 Computer Science Principles on the Prairie (CSPoP) program

The UNK Computer Science and Information Technology (CSIT) Department received funding in 2016 from the Google CS4HS grant program for the **Computer Science Principles on the Prairie (CSPoP)** program [2]. This was a year-long program that consisted of a face-to-face workshop, online coursework, and on-going community of practice activities to introduce K-12 teachers to computational thinking and the computer science "big ideas", as outlined in the AP Computer Science Principles curriculum [3], K-12 CS Framework [4], the CSTA CS Standards [5], and code.org [6], It was completed in active partnership with the State of Nebraska Department of Education, Nebraska Career Education (NCE).

**Program Rationale & Goals**

In Nebraska, there are no CS/IT state graduation requirements. Many rural Nebraska schools lack qualified teachers that can teach CS/IT classes. Only 10% of the high schools in Nebraska have some CS/IT graduation requirements. While Nebraska as a supplemental IT endorsement, teachers who teach CS/IT courses are not required to have this endorsement. Based on data from the NE Department of Education, out of the 63 high schools and 81 middle schools in Central Nebraska, only nine schools offer some IT-related course. Six offer a programming course; four offer a networking systems course; one offers a PC support course; and one offers an IT support course. Some of these schools are quite small (such as Plainview HS), indicating that it can be done by schools of all sizes. Schools in Western and Northern Nebraska are even more rural than Central Nebraska, and offer even fewer CS/IT offerings.

Teachers who currently teach IT courses in Nebraska usually attend NDE Business, Marketing & IT (BMIT) [7] workshops, for professional development and collaboration opportunities. A survey of 138 teachers who attended the 2015 NDE Business, Marketing & IT (BMIT) workshops self-reported that 57% of their high schools require an IT course for graduation, and 20% require two IT courses. The survey also found that coding is taught in 64% of self-reported school districts, but the survey found that schools did this in various ways, and with various amount of time dedicated to coding. Only 13% of the BMIT teachers reported that their schools offer a programming course.  It was encouraging to see these teachers interested in learning how to teach computer science, as the survey found that 63% of the BMIT workshop survey respondents wanted professional development in computer science/coding.

This program is based on the idea that all teachers are needed to be ambassadors for increasing CS/IT concepts into our educational system. In small rural schools that often do not have separate CS/IT teachers, it is imperative to work with any teacher interested in doing this – whether they are a 2nd grade teacher or a high school English teacher.

The ultimate goals of the CSPoP project were to tackle the IT workforce deficit in Greater Nebraska and make young people in rural Nebraska more IT-career ready, by helping interested rural Nebraska teachers develop CS/IT curriculum and helping them start on the path towards becoming certified to teach CS/IT courses, including AP and dual enrollment courses. The program also aimed to connect teachers (especially rural teachers) with each other and with computer science experts at the university.

By educating and providing support to the teachers who participate in this program, they are starting on the path towards becoming certified to teach CS/IT courses, including AP and dual enrollment courses. According to the Study of Dual Enrollment and Career Academies in Nebraska [8], "Nebraska ranks 49th out of 50 states in students, with only 12 percent of its high school seniors taking an AP exam as of 2011. Of those test-takers, 7.4 percent scored a 3 or above, which ranks Nebraska 47th." In Nebraska, students (and parents) prefer students to complete dual enrollment courses over AP courses, if they are available. "AP courses are high school courses taught at college rigor, whereas dual-enrollment courses are college courses, typically with identical syllabi, assessments and instructor qualifications updated on the college campus. To earn college credit in an AP course, a student must take and perform well on a single, end-of course examination, which colleges and universities can use to decide whether to offer credit for qualified scores related to the AP examination" [8]. By funding the first course of the IT teaching endorsement, it is likely that many of the stipend recipients will continue toward certification completion and towards earning their MSEd degree, which will also enable them to offer AP or dual enrollment CS/IT courses.

The CSPoP learning objectives were designed to help participating teachers to:
- Develop an information technology/ computer science (CS) curriculum for his/her age-level interest.
- Understand and apply computational thinking practices to curricular development for his/her age-level interest.
- Develop instructional strategies that create authentic and meaningful learning experiences, using the big seven computer science ideas (creativity, abstraction, data and information, algorithms, programming, the internet, and global impact) for his/her age-level interest.
- Discuss computational thinking and the big seven idea concepts with other teachers and administrators.

**Program Recruiting**

Information about this CS4HS program was provided via email to the Nebraska Educational Service Units (ESUs), the Nebraska Future Business Leaders of America (FBLA) Directors, and the Association for Career and Technical Education members. It was announced through the Nebraska Career Education newsletter, listserv, and website, along with a dedicated page on the UNK Department of Computer Science and Information Technology (CSIT) website.

**Professional Development Activities**

The program started with a face-to-face workshop offered as a pre-conference workshop at the Nebraska Career Education Conference (nceconference.com) [9] in June 2016. It introduced teachers to app development (using MIT App Inventor [10]), computational thinking, and several of the computer science big ideas [3]. This project provided funding for up to 30 teachers to attend this workshop.

The workshop was the initial meeting for a fully online seven-week graduate-level course offered through UNK. The course is the first course in the UNK IT supplemental endorsement program [11] for teaching CS/IT in Nebraska. This project funded up to 15 teachers who completed the workshop to complete the graduate-level online course over the summer.

The online graduate-level course, entitled IT Teaching Methods, requires reading and assignments on CS/IT curricular issues, computational thinking concepts, and the computer science big seven ideas, especially Creativity, Abstraction, Data and Information, Algorithms, and Programming (using App Inventor and Scratch [12]). The online course is set up in four modules that cover CS curricular development, computational thinking, the computer science big seven ideas, and programming practice. The teachers engaged with each other in multiple discussions on each of these topics. The teachers completed several programming assignments (using either Scratch or App Inventor), including projects that they design to be used within their own classroom, and to develop a curricular unit for their classroom. Because the teachers design the curriculum and programming projects for their own classrooms, the likelihood of implementation success is high.

Additionally, throughout the course, the teachers were required to engage in discussions on what computing courses currently are offered at their school; what CS/IT graduation requirements they believe schools should have; how ideas from this course could be incorporated into existing courses across the curriculum; what it would take for their school to develop computer science courses; and what it would take for them to become certified to teach the CS/IT courses, including the AP Computer Science Principles course and dual enrollment introductory to computer science university-level courses.

There were numerous teaching resources produced from this program that can be scaled and shared. These include the instructional material for the online course and the curricular units and programming projects that the teachers created as part of this project. Additionally, as the teachers returned to their classrooms, the projects that their students created could also be shared.

### Community of Practice Activities

During the school year, participant engagement will be facilitated by using a website of resources, and a Google community. The professional community allowed them to meet online throughout the year to exchange ideas and stories about what worked, etc. The professional community includes the teachers who participated in this CS4HS program as well as all teachers enrolled in the UNK IT Supplemental Endorsement program, and teachers who have previously completed the IT Teaching Methods course.

The summer activities that the teachers participating in this program completed, provided them with numerous required discussions and reviews of each other's work. These activities helped establish a trusting environment that encouraged open dialogue and reflective discourse.

## 3  Results

This program is designed to help rural Nebraska schools gain qualified CS teachers. The teachers who participated in this program did not have any formal CS education. However, the participants were all certified teachers who are competent at teaching at their grade level, and who sought professional development to teach CS/IT concepts. This program's success is based on the participants' desire to learn, along with the program administrator's experience of teaching the IT Teaching Methods course for the past four summers and the wealth of online materials available for teachers to learn how to teach CS/IT. It also helped that the teachers knew that they were designing curriculum and programming projects for their own classrooms.

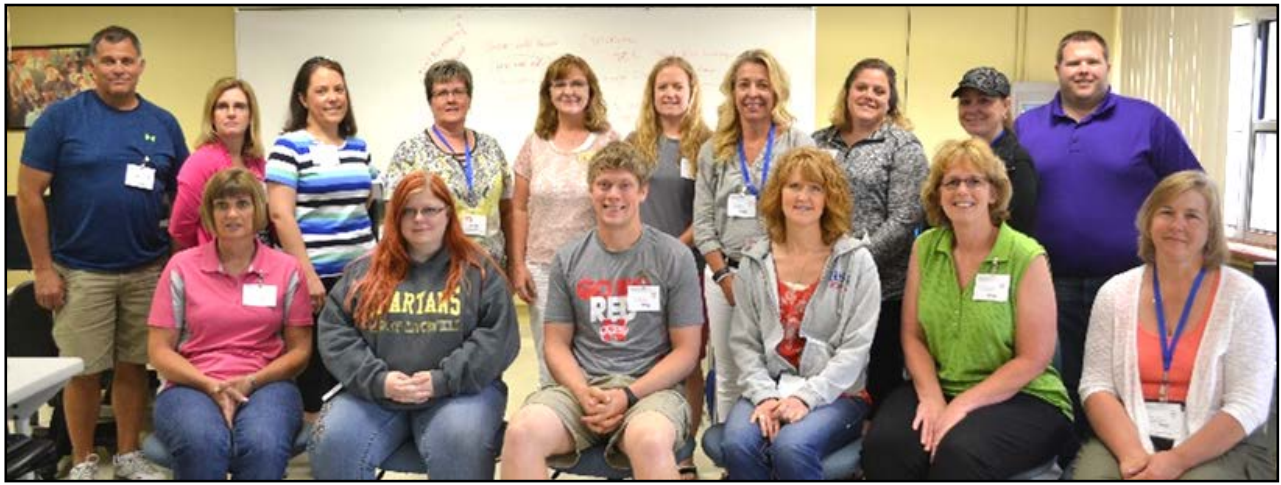

*Figure 2. Participants in the UNK 2016 CSPoP Mobile App Workshop*

Fifteen teachers participated in the NCE mobile app workshop and 15 teachers completed the CSIT 834P course. Several of these teachers indicated that they have implemented the curriculum they developed into their courses. Three of the participants were Math educators, one was a school tech coordinator, and the rest were business/IT teachers. Two teachers were from the Omaha school district; one was from Lincoln, and the rest were from rural areas.

Comments from the teachers include:
- I really enjoyed the class last summer and have used a lot of the material for the class that I am currently teaching.
- The opportunity to take the course free of charge was a huge plus for me. All of the learning in the course has been a huge benefit to me in the classroom.
- I really enjoyed this course. The instructor had us do a variety of different computer science activities that introduced what computer science entailed and what computational thinking was. I was able to come up with numerous ways to incorporate these in my math class. After taking this class, I am now wanting to become highly qualified in computer science so I can teach computer science classes at my school. This was the first computer science class I have ever taken and I loved it.
- I just wanted to say, I LOVE the App Inventor we've been introduced to. I love just tinkering on there, as well as following the tutorials to learn new things, and even using Google to figure out how to do something I want done. One of my favorite apps we've created and enhanced has been the PONG game. PONG is such a classic, and I had a lot of fun recreating it. My kids and I still load it and play the enhancement on my phone sometimes. It isn't very educational, necessarily, but the process of creation was!

We have 25 teachers in our Google Community. About 1/3 of the teachers have utilized the community of practice web resources throughout the year. We offered to host monthly video conference meetings, however, there was no interest, due to time constraints. One teacher commented, *"As far as the getting together or video conference piece, I am personally way too busy. I coach 3 different sports so between that and family I don't know that I could commit to doing any type of meeting."*

## 4  Lessons Learned & Future Work

During the first year of this CSPoP program, there were no face-to-face meetings during the school year. Teachers did not have time for video conferencing. Because of the relatively low usage of the community of practice web resources, we believe that we need to find better ways to fully engage and support the teachers through the year. Most of the teachers in the CSPoP program are the only teacher in their school district introducing any CS or computational thinking to the students. In these rural areas, these teachers are sometimes the only teacher within hundreds of miles willing to add CS concepts to their curriculum. It can be very scary and lonely endeavor. Having opportunities for teachers to meet each other in person is critical to the success of this program.

For future years, we plan to add fall and spring workshops to the CSPoP program. These will provide more opportunities for teachers to learn and interact together, and add more professional development opportunities for them.

We plan to do this by offering three hands-on face-to-face workshops during the fall at various Nebraska Educational Service Unit (ESU) [13] locations in central and western Nebraska. All of the school districts in ESUs 9-16 will be invited to send a teacher and 4-6 of their students to the workshop. We will also invite the teachers who participated in the CSPoP summer activities.

We expect to have 40-50 teachers in total participate in these workshops. We will focus on teachers in grades 6-9 and their students. The agenda for these events will include coverage of computational thinking concepts and the computer science big seven ideas using several CS unplugged activities and presentations. Nebraska Career Education (NCE) is creating "CS in a Box" resources that will be used at these workshops, and for the teachers to take home. These are based on the CS Unplugged [14] and CS in a Box [15] resources. During the lunch hour, the teachers will learn about resources for integrating CS into curriculum, while the students will attend a panel of professionals who will discuss CS/IT careers and the variety of career fields one could explore with CS.

By including some students in the fall workshops, the teachers are more likely to see the immediate benefit of incorporating computational thinking into their curriculum, and be more motivated to do so. Also, by bringing together teachers and students from several schools, they will have opportunities to share and learn from each other – and see that they are not alone as they implement CS concepts into their curriculum.

We also plan to offer face-to-face hands-on workshops in spring at the UNK TechEdge Conference for teachers [16]. We expect 60 teachers in total participate in these workshops. One workshop will be a follow-up for the teachers who attended the fall workshops and will discuss curricular ideas; and anther workshop will be designed for teachers who have no experience or background in CS/IT to introduce them to computer science teaching resources. The spring workshop will reinforce the fall workshop activities for the teachers and provide another opportunity for them to share and learn from one another.

## 5  Conclusion

Helping teachers in rural underserved areas develop CS curriculum and starting them on the path to becoming certified to teach CS courses is critical to making young people in underserved areas prepared for CS education at the university level, and ultimately helping tackle the IT workforce deficit in the United States. The Google CS4HS grant program is designed to facilitate the development of professional development and communities of practice for teachers to successfully develop and deliver CS curriculum in K-12 schools. The grant program is set up to allow projects to focus on local needs and resources.

This paper discussed a successful implementation of a Google CS4HS grant to a rural underserved area, as well as lessons learned through the implementation of the program. Key elements in the implementation included a face-to-face hands-on workshop, followed by a seven week graduate-level online summer course for the teachers to learn and develop curriculum that covers the CS concepts they will be teaching. The teachers enjoyed the course and incorporated CS curriculum into their classes.

The teachers were supported with an online community of practice for the year as they implemented the curriculum. A key lesson learned was that we needed to provide more face-to-face professional development opportunities during the school year. This will also improve the community of practice that supports teacher learning throughout the school year.

## Acknowledgments

The author is grateful to Google CS4HS for funding the CSPoP project, and remains indebted to the many high school teachers who participated in this project.

## References

[1] Google CS4HS program, retrieved March 16, 2017 from https://www.cs4hs.com/.
[2] UNK CSIT Google CS4HS Grant, retrieved March 16, 2017 from http://www.unk.edu/academics/csit/GoogleCS4HSGrant.php.
[3] AP Computer Science Principles curriculum, retrieved March 16, 2017 from https://secure-media.collegeboard.org/digitalServices/pdf/ap/ap-computer-science-principles-course-and-exam-description.pdf.
[4] K–12 Computer Science Framework, retrieved March 16, 2017 from https://k12cs.org/.
[5] CS Teachers Association, retrieved March 16, 2017 from http://www.csteachers.org/.
[6] Code.org, retrieved March 16, 2017 from code.org.
[7] Nebraska Department of Education Business, Marketing and Information Technology Career Field Cluster, retrieved March 16, 2017 from https://www.education.ne.gov/bmit/.
[8] Nebraska's Coordinating Commission for Postsecondary Education (CCPE) Dual Enrollment Report, retrieved March 16, 2017 from ccpe.nebraska.gov/sites/ccpe.nebraska.gov/files/doc/Dual_Enrollment_Report.pdf.
[9] Nebraska Career Education Conference retrieved March 16, 2017 from http://nceconference.com/.
[10] MIT App Inventor, retrieved March 16, 2017 from http://appinventor.mit.edu/explore/.
[11] UNK MSEd IT Supplemental Endorsement, retrieved March 16, 2017 from www.unk.edu/academics/csit/programs/msed-it-supplemental-endorsement.php.
[12] Scratch, retrieved March 16, 2017 from https://scratch.mit.edu/.
[13] Nebraska Education Service Units retrieved March 16, 2017 from http://www.esucc.org/nebraska-esus.
[14] CS Unplugged, retrieved March 16, 2017 from http://csunplugged.org/.
[15] CS in a Box, retrieved March 16, 2017 from https://www.ncwit.org/resources/computer-science-box-unplug-your-curriculum.
[16] UNK TechEdge Conference for teachers retrieved March 16, 2017 from http://www.unk.edu/academics/coe/unk-tech-edge-conference.php.